\documentclass[]{emulateapj}

\usepackage{longtable}
\usepackage{graphicx}
\usepackage{amssymb}

\shorttitle{ASASSN-15no}
\shortauthors{Benetti et al.}

\newcommand{\msun}{\,M$_{\odot}$}

\newcommand{\kms}{\,km~s$^{-1}$}

\newcommand{\ang}{\,$\rm{\AA}$}

\newcommand{\Ha}{\,H$\alpha$}

\newcommand{\hbeta}{\,H$\beta$}

\newcommand{\mcento}{mag\,(100d)$^{-1}$}

\newcommand{\ass}{\,ASASSN-15no}

\begin{document}

\title{ASASSN-15no: The Supernova that plays hide-and-seek}
%\title{ASASSN-15no: An energetic stripped envelope supernova with some residual Hydrogen in the ejecta}
\author{
S.~Benetti\altaffilmark{1},
L. Zampieri\altaffilmark{1},
A.~Pastorello\altaffilmark{1},
E.~Cappellaro\altaffilmark{1},
M. L. Pumo\altaffilmark{2,1,3},
N.~Elias-Rosa\altaffilmark{1},
P.~Ochner\altaffilmark{1},
G.~Terreran\altaffilmark{1,4,5},
L.~Tomasella\altaffilmark{1},
S.~Taubenberger\altaffilmark{6,7},
M.~Turatto\altaffilmark{1},
A.~Morales-Garoffolo\altaffilmark{8},
A.~Harutyunyan\altaffilmark{9},
L.~Tartaglia\altaffilmark{10,11}
}

\altaffiltext{1}{INAF - Osservatorio Astronomico di Padova, Vicolo dell'Osservatorio 5, 35122 Padova, Italy}
\altaffiltext{2}{Universit\'a degli Studi di Catania, DIEEI and DFA, Via Santa Sofia 64, 95123 Catania, Italy}
\altaffiltext{3}{INFN-Laboratori Nazionali del Sud, Via Santa Sofia 62, 95123 Catania, Italy}
\altaffiltext{4}{Astrophysics Research Centre, School of Mathematics and Physics, QueenÕs University Belfast, Belfast BT7 1NN, UK}
\altaffiltext{5}{UniPd}
\altaffiltext{6}{Max-Planck-Institut f\"ur Astrophysik, Karl-Schwarzschild-Str. 1, D-85748 Garching, Germany}
\altaffiltext{7}{European Southern Observatory, Karl-Schwarzschild-Str. 2, D-85748 Garching, Germany}
\altaffiltext{8}{Department of Applied Physics, University of C\'adiz, Campus of Puerto Real, 11510 C\'adiz, Spain}
\altaffiltext{9}{INAF-Fundacion Galileo Galilei, Rambla Jose Ana Fernandez Perez 7, E-38712 Brena Baja, Spain}
\altaffiltext{10}{Department of Astronomy and Steward Observatory, University of Arizona, 933 N Cherry Ave, Tucson, AZ 85719, USA}
\altaffiltext{11}{Department of Physics, University of California, 1 Shields Ave, Davis, CA 95616, USA}

\begin{abstract}

We report the results of our follow-up campaign of the peculiar supernova ASASSN-15no, based on optical data covering $\sim300$~ days of its evolution.
Initially the spectra show a pure blackbody continuum. After few days, the HeI $\lambda \lambda$ 5876 transition appears with a P-Cygni profile and an expansion velocity of about 8700 \kms. Fifty days after maximum, the spectrum shows signs typically seen in interacting supernovae. A broad (FWHM$\sim8000$ \kms) \Ha~ becomes more prominent with time until $\sim150$ days after maximum and quickly declines later on. At these phases \Ha~starts to show an intermediate component, which together with the blue pseudo-continuum are clues that the ejecta begin to interact with the CSM.
The spectra at the latest phases look very similar to the nebular spectra of stripped-envelope SNe. The early part (the first 40 days after maximum) of the bolometric curve, which peaks at a luminosity intermediate between normal and superluminous supernovae, is well reproduced by  a model in which the energy budget is essentially coming from ejecta recombination and $^{56}$Ni decay. From the model we infer a mass of the ejecta M$_{ej}=2.6$ \msun; an initial radius of the photosphere R$_0 = 2.1\times 10^{14}$ cm; and an explosion energy E$_{expl} = 0.8$ $\times 10^{51}$ erg. A possible scenario involves a massive and extended H-poor shell lost by the progenitor star a few years before explosion. The shell is hit, heated and accelerated by the supernova ejecta. The accelerated shell+ejecta rapidly dilutes, unveiling the unperturbed supernova spectrum below. The outer ejecta start to interact with a H-poor external CSM lost by the progenitor system about 9 -{}- 90 years before the explosion.
\end{abstract}


\section{Introduction} \label{intro}
The final core-collapse (CC) of massive stars can produce a wide variety of events in terms of luminosity: luminous stripped-envelope (SE) supernovae (SNe) 
with $M$ brighter than $-18.5$ mag \citep{maz13}, very weak SNe \citep[e.g., fainter than $\sim -14$ mag;][]{pas04,bot09,spi14} or even failed SNe - when 
the star directly collapses onto a black hole without significant mass ejection \citep{heg03,ada16,rey15}.

Another class of very luminous transients, named super-luminous supernovae \citep[SLSNe, with $M< -21$ mag;][]{gal12} has been identified in recent years. 
These can occasionally reach a peak absolute magnitude exceeding $-23.5$ mag \citep{don16,ben14}. SLSNe sometimes show spectral signatures of H in their spectra (SLSNe-II), but in most cases their spectra are H-free (SLSNe-I) and they resemble those of stripped-envelope SNe at late phases \citep{pas10,qui11}. The mechanism that supports the tremendous energy released in SLSNe is not yet understood, but the most promising central engines are: a) either a spinning-down magnetar \citep[e.g.][]{kas10,woo10,ins13} or an accreting black hole \citep{dex13}; b) the shock of the ejecta into a circumstellar shell (which can be H rich/poor) expelled by the progenitor star some time before the explosion \citep[e.g.][]{bli10,che11, whe17}; and c) the radioactive decay of a large amount of nickel produced in a pair-instability explosion (\citet[e.g.][]{gal09} but see \citet{mor10,you10,nic13}).

The luminosity gap between normal and super-luminous SNe is mostly populated by objects that show signs of interaction between fast expanding ejecta and a pre-existing circumstellar medium (CSM). The interacting system may be H rich, producing Type IIn SNe \citep{sch90,fil97,tur07} or He rich, producing the so called Type Ibn SNe \citep[][and references therein]{pas16}. 

The events populating this luminosity gaps can be very different as testified by the very recent supernova SN~2017dio \citep{kun17}, showing early spectra typical of a SE-SN, whose ejecta soon collide with a H-rich CSM, showing for the first time the transition from a Type Ic to a bright Type IIn in just few weeks.

In these objects the overall luminosity output include input from the ejected $^{56}$Ni \citep[which was shown to scale with the kinetic energy,][]{ter17}, plus the conversion part of the kinetic energy of the ejecta into radiation in the interaction process, which depends on the physical properties of the CSM shell (such as density and radius). Assuming that the photosphere radiates as a black-body, the luminosity increases with the radius, and/or with the photospheric temperature.

In the following sections, we will discuss the physical properties of \ass, an object in the luminosity gap between normal SNe and SLSNe. 

Our object was discovered by the ASAS-SN survey \citep{sha14} on 2015 August 3.26 UT \citep[][]{hol15,hol17}, and initially reported as an old Type Ic SN by \citet{bal15}.
Actually, as we will show in this paper, the spectrum eventually developed broad H lines  later on, that classifies the event as a Type II SN, caught at very early age.
The redshift of the faint ($g=18.57$ from SDSS DR10), cigar shaped, host galaxy (SDSS J153825.20+465404.1) measured from unresolved emission lines in our Gran Telescopio Canarias (GTC) spectra (see Section \ref{spectroscopic}), is $z=0.03638\pm0.00008$. Assuming $H_0$=73~km~s$^{-1}$~Mpc$^{-1}$ \citep{rie16} and a flat cosmology with $\Omega_{m}=0.31$ \citep{pla15}, we obtain a luminosity distance $D_L=153.5\,\rm{Mpc}$, hence a distance modulus of $\mu=35.93~\rm{mag}$.
For the foreground Galactic extinction, we adopt $A(V)=0.045\,\rm{mag}$ \citep{2011ApJ...737..103S}. Instead, since there is no evidence in the spectra of \ass~for the presence of a narrow \ion{Na}{1D} absorption feature at the recessional velocity of the parent galaxy, we assume negligible host galaxy extinction.

\section{Observations and data reduction} \label{obsData}
The follow-up campaign of \ass~was carried out using several facilities, namely the $1.82\,\rm{m}$ Copernico Telescope equipped with AFOSC (located at Mount Ekar, Asiago, Italy); the $10.4\,\rm{m}$ GTC with OSIRIS; the $2.56\,\rm{~m}$ Nordic Optical Telescope (NOT) with ALFOSC; Telescopio Nazionale Galileo (TNG) with Dolores (these telescopes are located at the Observatorio del Roque de los Muchachos, La Palma, Spain); and the Telescopi Joan Or\`o with MEIA2 located at Mount Montsec, Catalonia, Spain. Additional $V$ photometry has been supplied by the ASAS-SN telescopes.

\begin{figure} 
\begin{center}
\includegraphics[width=1.05\linewidth]{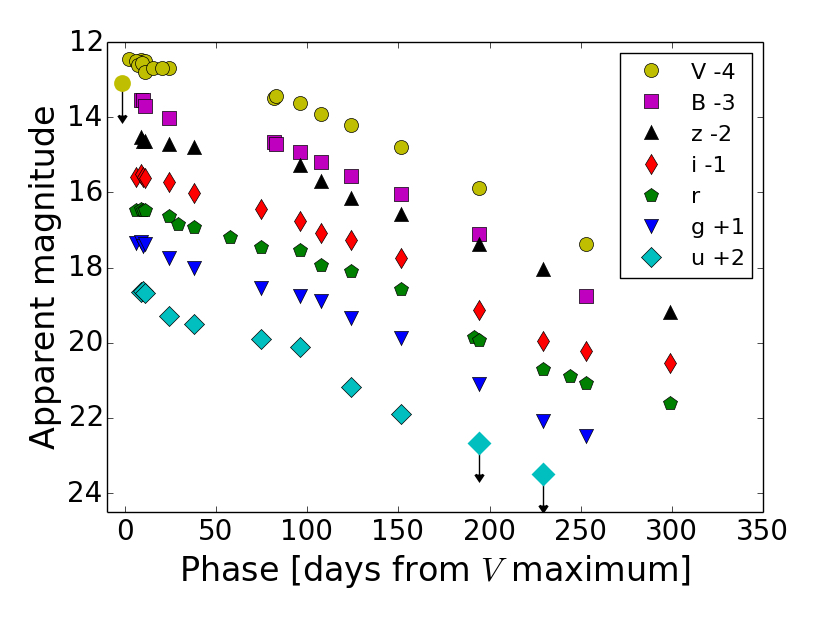}
\caption{{\it BV} (Vega) and {\it ugriz} (AB mag) light curves of \ass. The phase is in the observer's frame.} \label{photometry}
\end{center}
\end{figure}

Spectroscopic and photometric data were pre-processed in the standard manner (performing overscan, bias and flat-field corrections) using \textsc{iraf}\footnote{\url{http://iraf.noao.edu/}} tasks.
For the photometric measurements, the SNOoPY \citep{snoopyref} package was used, which allowed to extract the magnitude of the SN with the point-spread-function (PSF) fitting technique \citep{stet87} for the {\it BVRI} frames, and the template subtraction technique for the {\it ugriz} frames (no deep template frames were available for BVRI bands). Indeed, {\it ugriz}~template frames from the SDSS survey were obtained on June 2012, three years before the SN break out.
The SN magnitudes were then calibrated with reference to the magnitudes of field stars retrieved from the SDSS catalog (DR9). 
For the {\it BVRI} filters, we converted the SDSS catalog magnitudes to Johnson/Cousins, following \cite{2008AJ....135..264C}. The magnitudes of the Johnson/Cousins bands are thus given in the Vega system, while the SDSS ones are calibrated in the AB system. 

Spectra of comparison lamps obtained with the same instrumental set-up were used for the wavelength calibration and standard stars observed in the same nights were used as reference for the flux calibration.
The wavelength calibration was checked using night sky lines.
The flux calibration was verified against the broad band photometry, and a scaling factor was applied when necessary.
Finally, we corrected the spectra of \ass~for the Galactic foreground extinction and for the redshift of the host galaxy.

\begin{deluxetable*}{@{}crcccc@{}}
\tablecaption{Log of the spectroscopic observations of \ass.}
\tablehead{\colhead{Date} & \colhead{Rest phase} & \colhead{Instrumental set-up} & \colhead{Grism or grating} & \colhead{Spectral range} & \colhead{Resolution} \\ 
\colhead{} & \colhead{} & \colhead{} & \colhead{} & \colhead{(\ang)} & \colhead{(\ang)} }
\startdata
20150806 & $+6$ & Ekar182$+$AFOSC & gm4 & $3400-8250$ & 13.5 \\
20150809 & $+9$ & Ekar182$+$AFOSC & gm4 & $3400-8250$ & 13.5 \\
20150810 & $+10$ & Ekar182$+$AFOSC & gm4\,/VPH6 & $3400-9280$ & 14 \\
20150811 & $+11$ & Ekar182$+$AFOSC & gm4 & $3400-8250$ & 13.5 \\
20150812 & $+12$ & Ekar182$+$AFOSC & gm4 & $3400-8250$ & 13.5 \\
20150817 &$+16$& TNG$+$LRS            & LRB\,/R & $3230-9650$ & 15 \\
20150824 & $+23$& NOT$+$ALFOSC & gm4 & $3450-9020$ & 18 \\
20150829 & $+28$& GTC$+$OSIRIS & R1000B\,/R & $3640-9390$ &  7 \\
20150830 &$+29$& TNG$+$LRS            & LRB & $3300-8080$ & 11.0 \\
20150907 & $+37$ & Ekar182$+$AFOSC & gm4\,/VPH6 & $3500-9300$ & 14 \\
20150927 & $+56$& GTC$+$OSIRIS & R1000B\,/R & $3640-10390$ &  7 \\
20151014 &$+72$& TNG$+$LRS            & LRB\,/R & $3400-9640$ & 16 \\
20151104 & $+92$ & Ekar182$+$AFOSC & gm4\,/VPH6 & $3500-9300$ & 14 \\
20151204 & $+121$ & Ekar182$+$AFOSC & gm4 & $3460-8240$ & 13.5 \\
20151229 & $+146$& NOT$+$ALFOSC & gm4 & $3350-9070$ & 18 \\
20160208& $+184$& GTC$+$OSIRIS & R1000B\,/R & $3630-10400$ &  7 \\
20160401& $+235$& GTC$+$OSIRIS & R1000B\,/R & $3640-10390$ &  7 \\
20160526& $+289$&GTC$+$OSIRIS & R1000R       & $5100-10390$ &  8 \\
\enddata
\tablecomments{The rest phases are relative to the $V$ band maximum (2015 August 1).
}
\label{specLog}
\end{deluxetable*}

\section{Photometric analysis}
The $ugriz$ and $BV$ light curves are shown in Figure~\ref{photometry}, while the magnitudes in all bands (those shown in Figure~\ref{photometry} plus $URI$) are reported in the online Tables \ref{sdss_phot} and \ref{vega_phot}. 

The pre-discovery limit and discovery $V$ magnitudes reported by \citet{hol15} have been revised using deeper, higher quality reference images, and are shown in Figure \ref{photometry}.  No deep pre-discovery observations are available for BRI bands. This allows to constrain the $V$-band maximum to 2015 August $1\pm 2$ d (MJD=57235) and to infer that the rise time, in this photometric band, could have been relatively short.

The $u$ to $z$ light curves have very similar shapes, showing relatively fast declines from the peaks lasting about 20-40 days (we do not have a dense temporal coverage of this phase). Later on, the light curves settle on a slower linear decline up to about 100 days after maximum, with rates only mildly depending on the pass-band (e.g. $1.15\pm0.06$ \mcento~ in $u$; $1.24\pm0.14$ \mcento~in $r$; and $0.75\pm0.04$ \mcento~in $z$). After 100 days, all light curves show a more rapid decline rate expecially in blue bands  (e.g. $>2.41$ \mcento~in $u$; $2.30\pm0.08$ \mcento~in $r$; and $1.98\pm0.10$ \mcento~in $z$). The latest $r,i$ and $z$ points at $\sim 300$ days suggest a possible flattening, but the large uncertainties in the measurements prevent us from a robust statement.

\section{Spectroscopic analysis} \label{spectroscopic}
\begin{figure*} 
\begin{center}
\includegraphics[width=0.9\linewidth]{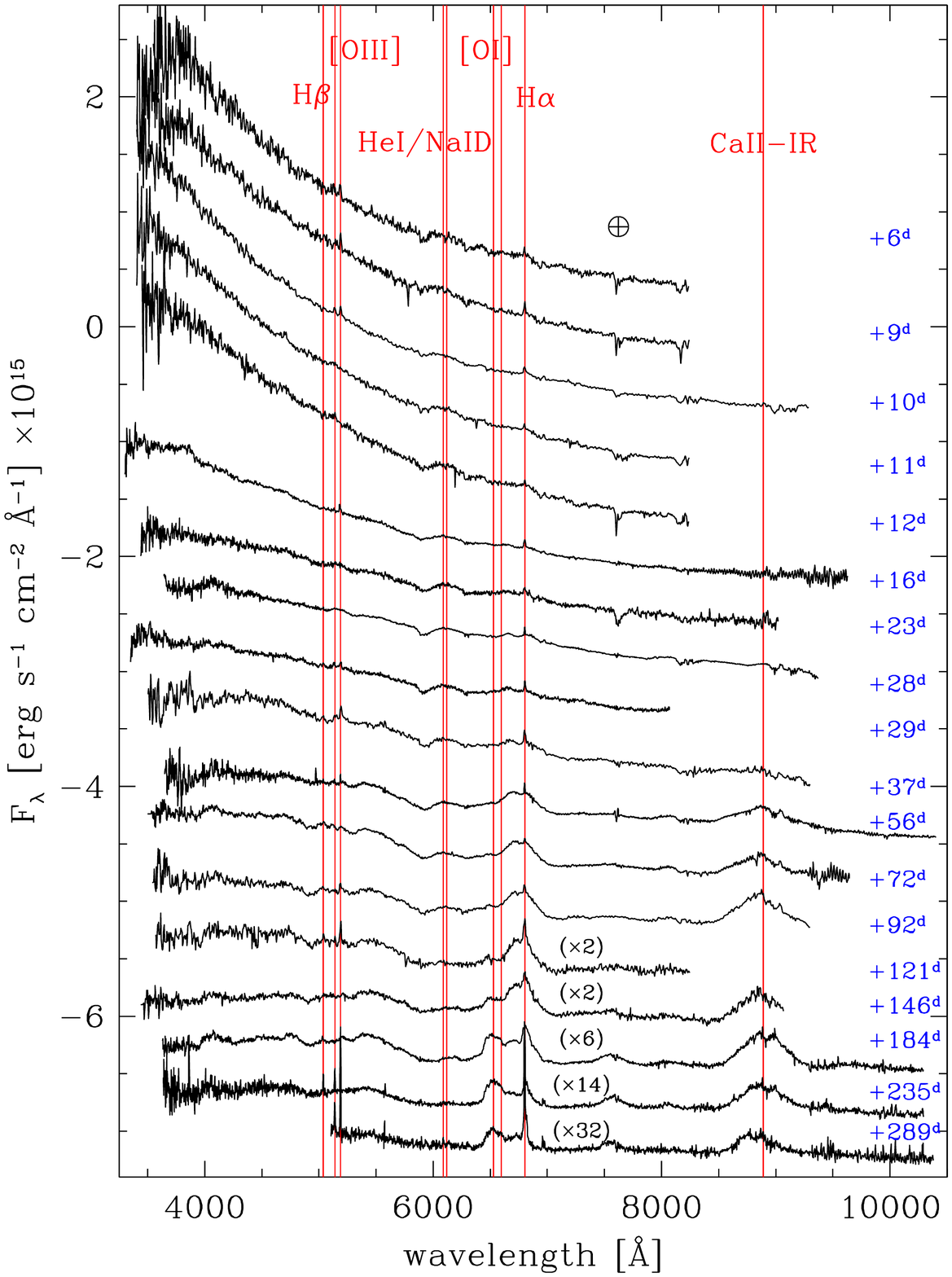}
\caption{Spectroscopic evolution of \ass. The wavelength scale is set in the observer's frame and are not corrected for the Galactic extinction. The ordinate refers to the first spectrum, and the others have been arbitrarily shifted downwards. For the sake of clarity, the flux counts in the last five spectra have been multiplied by the numbers given in parenthesis. Residuals from the atmospheric absorption corrections have been marked with the $\oplus$~symbol.} \label{spectroscopy}
\end{center}
\end{figure*}

The journal of the spectroscopic observations is reported in Table~\ref{specLog}, along with the information on the instrumental configurations used and spectral resolutions. The results of the spectroscopic follow-up campaign of \ass~are shown in Figure~\ref{spectroscopy}. The spectra are shown in the observer's frame and are not corrected for the Galactic extinction.

The spectra at early phases are characterised by a blue continuum.
The temperature (estimated through a black-body fit) initially very high, $13300\pm300\,\rm{K}$ at $+6$ d, rapidly drop to $8200\pm300\,\rm{K}$ at $+23$ d, and stays nearly constant hereafter (see Section \ref{temp-radius}). The first feature clearly seen in the spectra beginning at $+11$ d onwards is a line with a broad and shallow P-Cygni profile probably due to He~I 5876\AA. The expansion velocity deduced from the minimum of its absorption is about 8700 \kms. At the same time a broad (FWHM$\sim8000$ \kms) \Ha~emission begins to grow. On day $+37$, the spectrum starts to show a growing blue continuum component blue-ward of $\sim 5600$ \AA, which increases with time. This blue pseudo-continuum is often seen in interacting supernovae \citep{tur93,smi09} and is due to the blending of Fe line emissions. By day $+56$, the spectrum shows very broad \Ha~and CaII-IR emissions. The \Ha~profile is complex, consisting of a broad base with a half width at zero intensity of $\sim 23000$ \kms~for the blue wing and $\sim 13000$ \kms~for the red wing. An intermediate (FWHM $\sim 12000$ \kms) component, blue-shifted by $\sim -2000$ \kms~with respect to the \Ha~rest frame wavelength, is superimposed to the broad emission. Starting from day $+72$ a broad \hbeta~absorption begins to appear and by day $+92$ an expansion velocity of about 7400 \kms~is derived from its minimum.

The CaII-IR emission has instead a triangular shape, with a FWHM $\sim 15400$ \kms~centered at almost its rest-frame position (see Figure \ref{spectroscopy}). This feature remains visible until late phases, with an almost constant FWHM. However, in the last two spectra, the FWHM decreases to $\sim 13700$ \kms~at day $+235$, and $\sim 11800$ \kms~at day $+289$, caused by a fading of the red wing, and a simultaneous blueshift of the whole emission by $\sim 70$ \AA. This could be an indication of dust formation in the shocked ejecta. \Ha~shows a constant profile until $+146$ d after maximum, and then fades. In the last three spectra it appears as a relatively faint, broad and boxy emission blended with a now dominant [OI] doublet. The last spectra are reminiscent of those of some stripped envelope SNe, although contaminated by narrow emissions of the parent galaxy, (see Section \ref{spec_comp}). Besides the broad, shallow emissions, the spectra later than $\sim +50$ days show a persistent blue pseudo-continuum, typically seen in interacting SNe.

\section{Absolute magnitudes and Bolometric light curve of \ass} \label{bolo}
With the values adopted for distance and extinction in Section \ref{intro} \cite[using the][extinction law]{card}, we obtain $M_{\rm V} \sim -19.6\pm
0.3$ mag at maximum light. Similar magnitudes are inferred for the $r$ and $B$ bands at maximum. These values are similar to those derived for some interacting SNe, such as SN~1997cy \citep{tur00} and SN~1999E \citep{rig03}.

\begin{figure} 
\begin{center}
{\includegraphics[width=0.95\linewidth]{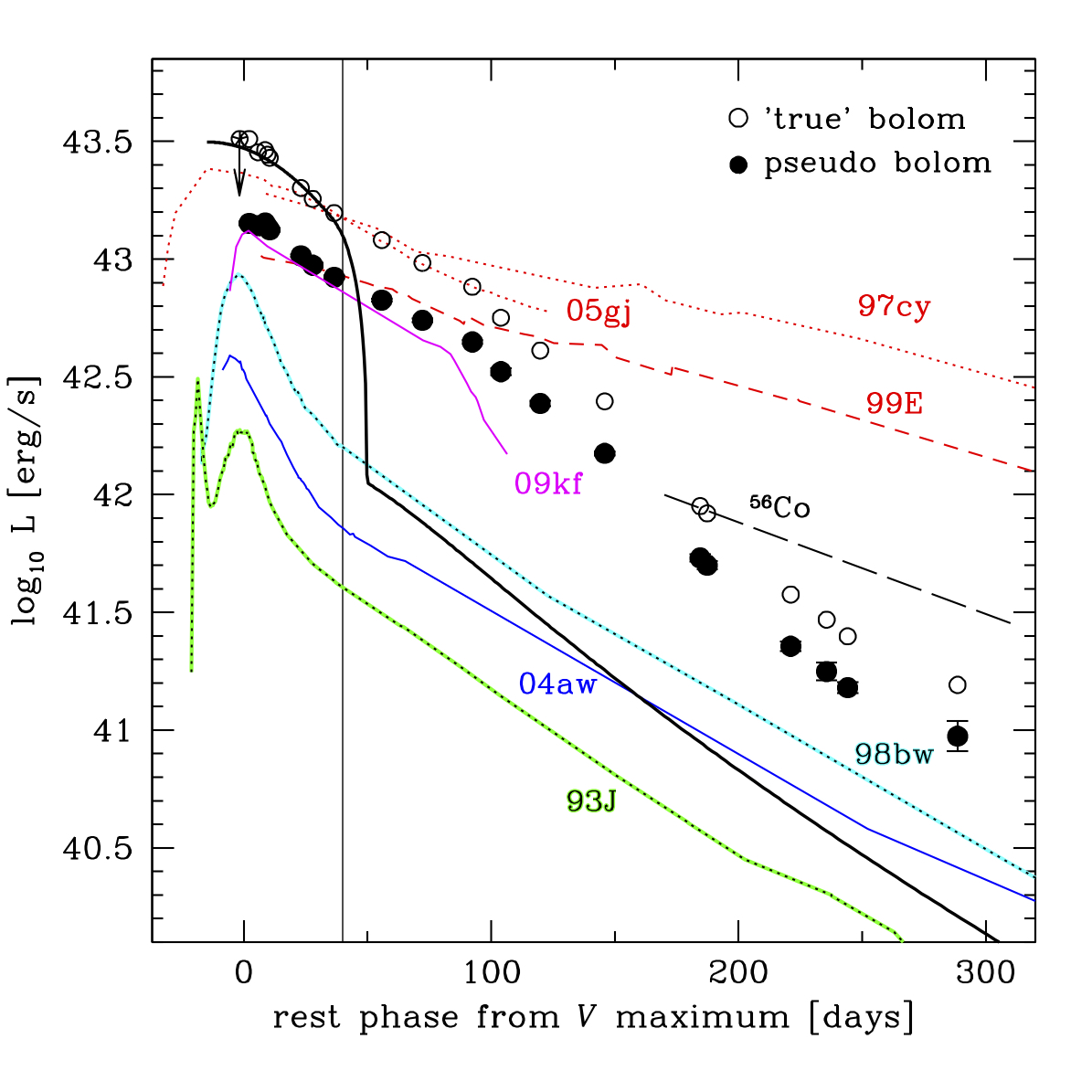}
\includegraphics[width=0.95\linewidth]{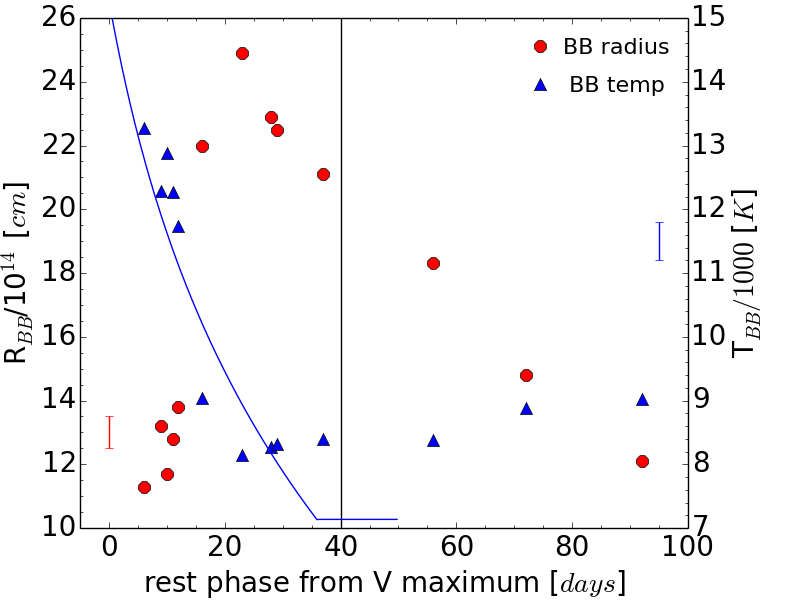}}
\caption{{\it Top}: Comparison of the pseudo-bolometric $uBgVriz$ light curve of \ass~(filled black circles), with those of a sample composed by some Type II and Type Ic SNe (integrated over the same wavelength range). The comparison sample includes (in order from high to low luminosities): SNe 2005gj \citep[IIn,][]{pri05,ald06}; 1997cy \citep[IIn,][]{tur00}; 2009kf \citep[II,][]{bot10}; 1999E \citep[IIn,][]{rig03}; 1998bw \citep[BL-Ic,][and references therein]{pat01}; 2004aw \citep[Ic,][]{tau06}, where the higher reddening discussed in \citet{ben06} has been adopted; and 1993J \citep[IIb,][and references therein]{bar95}.  The estimated full bolometric light curve of \ass, marked with open circles is also included. The initial limit of this curve (open-skeletal symbol) was calculated from the $V$-band limit (see Table \ref{vega_phot}) and an assumed temperature of $16200$K, as deduced from hydrodynamical modelling (cfr. Section \ref{modelling} and bottom panel). The black solid line is the bolometric curve of the best-fit model (see text for further details). The light vertical line marks the onset of the ejecta-CSM interaction (phase $\sim +40$ day).
For completeness the $^{56}$Co decay slope is also indicated dashed line). {\it Bottom}: \ass: evolution of the blackbody radius and the temperature. The average temperature error bar of $\pm 300$ K is shown on the right-hand side, while the mean error bar for the radius ($\pm 0.4\times 10^{14}$ cm) is shown on the left-hand side. The blue solid line is the photospheric temperature evolution of the best-fit model (see text for further details). The black vertical line marks the onset of the ejecta-CSM interaction (phase $\sim +40$ day).
}
\label{bol_radius}
\end{center}
\end{figure}

The pseudo-bolometric light curve of \ass~has been computed by integrating its multi-colour $uBgVriz$ photometry, neglecting any possible contribution in other spectral ranges. The UV contribution, however, is expected to be significant especially at early stages. For each epoch and filter, we derived the flux at the effective wavelength. We adopted as reference the epochs of the $r$-band measurements (except for the first two epochs, for which we used the $V$ band as reference). Missing fluxes in other filters were estimated through interpolation or, if necessary, by extrapolation from the closest available epoch assuming a constant colour with respect to adjacent bands. The flux measurements for all filters, corrected for extinction, give the spectral energy distribution at each epoch, which is integrated with the trapezoidal rule, assuming zero flux at the integration limits. The observed flux was then converted into luminosity. The resulting pseudo-bolometric light curve is shown in Figure \ref{bol_radius} (upper panel), along with those of a sample of SNe that includes SNe IIn with similar luminosity and spectral evolutions, the peculiar Type II SN~2009kf, and a few stripped envelope SNe, because of the similarity of the late time spectra of \ass~with this SN type (see Section \ref{spec_comp}).

A "true" bolometric light curve is also reported in Figure \ref{bol_radius}. To construct it, we had to estimate the bolometric correction for each phase. This was performed by fitting with a blackbody the $BgVriz$ SED (we didn't consider the $u$ band because of the heavy line blanketing affecting this spectral region) derived in epochs where the SN had multi-band detections, and then adding the missing flux, measured from 0 to $\infty$, to the optical luminosities. We remark that the BB temperatures derived from the photometric SED are in close agreement with those derived from the spectral fitting (see Section \ref{temp-radius}). Because of the blue pseudo continuum seen in later time spectra of \ass~due to ongoing interaction, the blackbody fit of the SED could led to an incorrect estimation of the temperature and accordingly to an overestimation of the bolometric correction. The "true" bolometric luminosity is used for the modelling discussed in Section \ref{modelling}.

The luminosity of \ass~is intermediate between those of the Type IIn SNe 2005gj, 1997cy,  and 1999E\footnote{These three SNe share similarities with the so-called Type Ia-CSM events. A prototypical event of this class is PTF11kx, which is believed to be a thermonuclear explosion in a H-rich CSM \citep{dil12}, but other events were proposed to belong to this group, including SNe 2002ic, 2005gj and  2012ca \citep{ham03,pri07,fox15}. We remark, however, that other authors suggested an alternative scenario with an energetic CC engine for these transients \citep{ben06,tru08,ins16}.}, and a few dex higher than that of the energetic stripped envelope (SE) SN 1998bw. The pseudo bolometric light curve shows a regular decline from the peak to $\sim 40$ days after maximum, followed by a slightly flatter decline up to about $+100$ d, similar to the comparison Type IIn SNe and SN~2009kf. However from $\ga 40$ d onwards \ass~shows a luminosity excess with respect to SN~2009kf. After $\ga +100$ d the decline rate of \ass~increases again. 

\begin{figure*} 
\begin{center}
\includegraphics[width=0.95\linewidth]{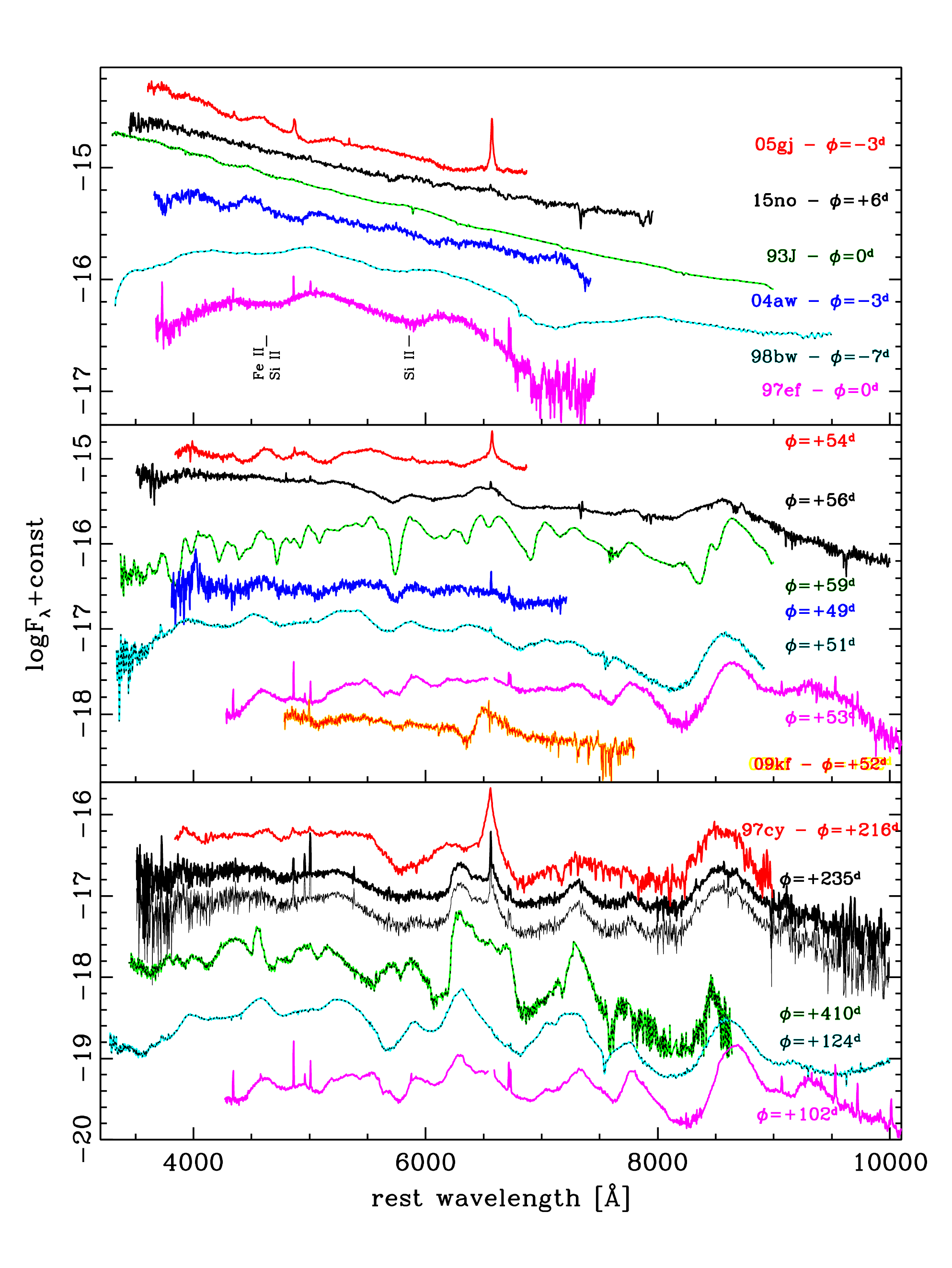}
\caption{Comparison of \ass~spectra at three key phases with those of reference objects. In the top panel (from top): SN 2005gj \citep[IIn,][]{pri05,ald06}, \ass, SNe 1993J \citep[IIb,][and references therein]{bar95}, 2004aw \citep[Ic,][]{tau06,mod14}, 1998bw \citep[BL-Ic,][]{pat01} and 1997ef \citep[BL-Ic,][]{iwa00}. The main Si II (6355 \AA~doublet and 5058 \AA), and Fe II (4924, 5018, 5169 \AA) absorption blends normally seen in Type Ic SNe are marked \citep[e.g.][]{maz00}. In the middle panel a +52 days spectrum of SN~2009kf \citep[II,][]{bot10} has been also included. In the bottom panel, SN 2005gj is replaced by SN 1997cy \citep[IIn,][]{tur00}. In this panel the \ass~spectrum is shown as observed (solid line) and with a BB continuum subtracted (lighter, lower line; see text for details).
}
\label{comp_spec}
\end{center}
\end{figure*}

\section{Discussion} \label{discussion}
\subsection{Photospheric temperature and radius evolution} \label{temp-radius}
The photometric and spectroscopic monitoring of the optical transient \ass~can be used to derive some physical properties of the transient, such as the evolution of the photospheric radius and temperature. We performed a blackbody fit to our early time spectra (phase $< + 100$ days), after correcting for the adopted reddening and redshift values (see Section \ref{intro}). The fits were done selecting the portions of the spectra free of emissions. The fit errors, depend on a number of parameters (including the wavelength extension of the spectra, the line contamination, the uncertainty in the flux calibration and the assumed extinction) and have been estimated to be of the order of 300 K. The temporal evolution of the temperature is shown in Figure \ref{bol_radius}, bottom panel (blue triangles). Adopting the true-bolometric light curve of Figure \ref{bol_radius} (upper panel), spherical symmetry and a filling factor of 1, we calculated the evolution of the blackbody radius (bottom panel in Figure \ref{bol_radius}), up to 100 days past $V$ maximum.

The photosphere around maximum is hot, with T$_{BB} \sim 13300$ K, then shows a steep decrease to about $8200$ K at a phase of +25 d.  Hereafter until day $+56$ the temperature remains constant, but then it smoothly increases up to $9000$~K. The late temperature behaviour (later than day $\sim +40$) is likely an artefact due to the contamination on the blue region of the spectra that we attribute to the on-set of ejecta-CSM collision. The photospheric radius, in the same time interval, steeply increases from $8\times 10^{14}$~cm to about $18\times 10^{14}$ cm (at a phase of $+25$ d), and then linearly decreases to about $10\times 10^{14}$ cm at about $+90$ -{}- $+100$ d (see Figure \ref{bol_radius}, bottom). It is interesting to note that the photospheric expansion rate in the first 25 d after maximum is about $6.64\times 10^{13}$ cm day$^{-1}$, which corresponds to about 7700 \kms. This is very similar to the expansion velocity measured from the He~I line visible in the spectra from $+ 7$ days after maximum (see Section \ref{spectroscopic}). We deduce that for about 20 days the photosphere was locked to the same layer in the expanding ejecta, because the ejecta are highly ionised and thus opaque, and only after 20 days when the temperature is low enough to start the recombination the photosphere moves inwards in mass coordinate.

\subsection{Modelling of the light curve} \label{modelling}
In order to better understand the early phases of the supernova and retrieve information about the physical properties of its progenitor at explosion, we compared the true-bolometric curve of ASASSN-15no with that predicted by the simulations described in \citet{zam03} and \citet{zam07}. This code solves the energy balance equation for ejecta of constant density in homologous expansion and yields the best fit model by simultaneously comparing the computed bolometric light curve, photospheric gas velocity and temperature with the corresponding observed quantities. 
Although it is less accurate than a full radiative-hydrodynamic treatment \citep[see e.g.][]{pum11,pum13}, it is able to provide reliable estimates of the physical conditions of the ejected gas and has been successfully applied to a number of SNe II (either to directly model the SN ejecta or to constrain the parameter space for fully hydrodynamic calculations; see e.g. \citet{zam03,pum17,zam17} and references therein). The early photospheric temperature evolution (lower panel of Figure \ref{bol_radius}) is well fitted during the first 40 days past maximum by the model assuming that the first energy input occurred 15 days before maximum light (with an uncertainty of about 5 days, found analysing the quality of the model best fits), and the early photospheric gas velocity is constant at a value close to 6300 \kms. This is 2000 \kms~ lower than the expansion velocity deduced from the minimum of the He I 5876 \AA~ transition seen in \ass~spectra ($<v_{HeI}>=8400\pm500 $ \kms~for phases $< +75$ days), and after applying a $v$(Fe~II 5169 \AA)/$v$(He~I 5876 \AA) correction derived from SN~1993J spectra, a prototype of a SE supernova. In fact, the photospheric velocity is usually best traced using Fe and/or Sc lines \citep{zam03,pum11} which, however, are not evident in the \ass~spectra.

As shown in Figure \ref{bol_radius}~the model well matches the bolometric light curve only for the first $\sim 40$ days after the $V$ maximum when the input energy is dominated by the recombination and radioactive decay. The physical parameters derived from the model are the following: mass of the ejecta, M$_{ej}=2.6$ \msun; initial radius of the photosphere, R$_0 = 2.1\times 10^{14}$ cm; explosion energy, E$_{expl} = 0.8\times 10^{51}$ erg. The $^{56}$Ni mass, M$_{Ni}$, and recombination temperature are fixed at  0.15 \msun~and 6000 K, respectively. The values of these parameters are reasonable for a typical CC explosion, except for the radius. The derived radius of $\sim 3000$ R$_{\odot}$ would be consistent only with a very extended supergiant star \citep[see e.g.][on Galactic red supergiant stars properties]{lev05}.\\
It is reassuring to note that the initial radius deduced from the model is in agreement with the measurements from spectrophotometry (see Figure \ref{bol_radius}, bottom panel).

The radius and, in general, the best-fitting model parameters (e.g. the low ejecta mass) are difficult to reconcile with the collapse of a single star, compared to a scenario involving a massive star in a binary (or multiple) system.

It is of interest to compare the physical parameters deduced for \ass~with those derived for SN~2009kf \citep{bot10}. In fact SN~2009kf share many physical properties (e.g. spectroscopic type, peak luminosity, early photometric evolution) with \ass. Modelling the physical parameters of SN~2009kf, \citet{utr10} found a somewhat lower initial radius (2000 R$_{\odot}$), but a much higher ejecta mass (28.1 \msun), explosion energy (22 foe) and $^{56}$Ni mass ($0.4$ \msun). The differences regarding ejecta mass and total energy are understood considering that in SN~2009kf the photospheric phase lasts 90 days and is only supported by the ejecta recombination and radioactive decay, while in \ass~lasts only $\sim 55$ days according to our scenario.

\subsection{Spectroscopic comparison with a sample of Core Collapse SNe} \label{spec_comp}
We already mentioned that the \ass~ light has a BB-like energy distribution at maximum, in analogy with other SNe (e.g. 2005gj and 1993J, see top panel of Figure \ref{comp_spec}). Instead, the near-maximum spectra of energetic SNe~Ic, despite of being dominated by a strong continuum, show also broad Fe~II and Si~II features.

The temperature of the emitting photosphere reaches its minimum value, when the radius of the photosphere is maximum, then the temperature remains almost constant, while the radius steadily decreases, and the recombination front recedes inward through the ejecta. However, the late time temperature evolution could be affected by the increasing strength of the blue pseudo-continuum, probably arising from a cool dense shell, the region where the ejecta-CSM interaction take place \citep{smi09}. 

Overall the observations suggest the following scenario. Initially \ass~ejecta is in free expansion, but $\sim 40$ d after maximum it starts to interact with a close-by CSM. This is confirmed by the spectroscopic comparison shown in the middle panel of Figure \ref{comp_spec}, where the spectrum of \ass~shows remarkable differences when compared with that of SN~2009kf. \Ha~is faint and broad, while in SN~2009kf it has a P-Cygni profile typically seen in non-interacting SNe. The \ass~spectrum shows a blue pseudo-continuum typically seen in the spectra of interacting SNe. Still, the maximum expansion velocity \citep[$\sim18000$ \kms, assuming an average Ca~II IR triplet wavelength of 8579.1 \AA,][]{maz05} derived from the broad Ca II IR features of \ass~and SN~1998bw is comparable.
After $+100$ d, the bolometric luminosity decline becomes steeper (upper panel of Figure \ref{bol_radius}) which probably correspond to a phase of fading interaction.
If, as assumed in our modelling, the explosion of \ass~has ejected a $^{56}$Ni mass of about 0.15\msun~similar to the amount of $^{56}$Ni ejected in other CC SNe, by $\sim +250$ days the contribution due to interaction reaches a minimum (see Figure \ref{bol_radius}, upper panel). By this time, the spectrum is thus a combination of the emission coming from the interaction, and the emission from the non-interacting SN. This is best seen in the bottom panel of Figure \ref{comp_spec}, where we show the \ass~spectrum as observed and with a subtracted BB contribution (assumed coming from the interaction and estimated from the flux difference from the \ass~bolometric luminosity with respect to that of the SE SNe, see top panel of Figure \ref{bol_radius}). The second (subtracted) spectrum shows typical nebular features of a SE SN with a weak \Ha. The simultaneous presence of interaction and ejecta features in the spectra likely indicates that the CSM is clumpy, or that its geometry and/or density structure are highly asymmetric. 

\subsection{6300-64 \AA - [OI] lines} \label{oi}
Following \citet{tau09}, we have de-blended the [OI] 6300-64 \AA~and \Ha~lines on the last three \ass~spectra. However, in \ass~spectra the de-blending is complicated by the presence of the broad \Ha.
The central position of the 6300 feature is similar to that observed in the SN Ib/c sample presented in Taubenberger et al, and it also experiences a similar red-ward shift with time shown by the SN Ib/c sample. The FWHM is large ($5000-7000$ \kms~at $+184$ days,  $5400-6500$ \kms~at $+235$ days and $5100-5900$ \kms~at $+289$ days, depending from the assumed \Ha~width) and very similar to the widths shown by the [OI] 6300 \AA~line in the more energetic SE-SNe \citep[see Figure 8 of ][]{tau09}.

\Ha, which is blended with the  [OI] doublet, has a composite profile. At $+186$ days, it consists  of a broad (FWHM $\sim 10000$ \kms) component, blue-shifted by $\sim 4000$ \kms~(similar to that observed in early spectra, see Section \ref{spectroscopic}), and an intermediate one (FWHM $\sim 3800$ \kms), redshifted by $\sim 1100$ \kms. At $+294$ d, while the broad component does not show a clear evolution with time, the intermediate one is found almost at the rest wavelength with a FWHM $\sim 1700$ \kms.

\subsection{The progenitor scenario} \label{progenitor}
In an attempt to address all the observed evidences, we sketch the following scenario. The stripped envelope C/O core of a massive star, undergoes a likely asymmetric (see below) explosion which is initially masked by an opaque H-poor/He-rich, detached shell having an inner radius of $\sim 3000$ R$_{\odot}$. 

Such a shell could have been lost by the SN progenitor during a common-envelope mass loss episode with a companion star, although the uncertainties on the common-envelope evolution prevent us from having precise information on this type of episodes \citep[see e.g.][and references therein]{iva13,tau17}.

Assuming an average expansion velocity of $\sim 15000$ \kms~for the ejecta and the initial radius deduced from the modelling ($\sim 3000$ R$_{\odot}$), we can estimate that the ejecta travelled about 1.7 days before hitting the shell (likely expanding at low velocity, $\la 100$ \kms). 
When the outer ejecta collide with the shell, a fraction of the kinetic energy of the ejecta heats the gas, and speeds up the shell itself. If the velocity deduced from the expanding photospheric radius (7700 \kms~cfr. Section \ref{bolo}) is in fact an indication that the ejecta plus swept-up shell is expanding with that velocity, this is about half of the velocity of the SN ejecta alone.

Assuming that the explosion of \ass~was similar to that of SN~1993J, whose observables were well fitted by models of \citet{woo94} and \citet{bli98}, with ejected mass as low as 1.54 \msun~(assuming binarity and the formation of a remnant of 1.55 \msun) we can estimate that the mass of CSM shell enclosing the \ass~progenitor was as high as $1$ \msun~(from our modelling the mass of the SN ejecta plus swept-up up material is $2.6$ \msun, see Sect. \ref{modelling}).

This scenario involving the ejecta-shell collision could be a scaled down (both in mass and radius of the shell) version of those proposed as a possible explanation for some SLSNe-II. As an example, in the H-rich SLSN CSS121015:004244+132827 \citep{ben14} the initial radius has been estimated to be $\la 2\times10^{15}$ cm and the shell mass to be $\sim 8.5$\msun, making this SLSN at least $1.0$ dex more luminous than \ass.

The ejecta plus swept-up shell material begin a mild interaction with an outer CSM about 40 days after maximum. This outer CSM has probably high density and is H-poor. We get to these conclusions because in the early spectra there is no sign of low velocity Balmer recombination lines. The faint narrow \Ha~line seen in early spectra mostly comes from an underlying H-II region which contaminates the spectra at all phases (see also the presence of the [S II] doublet in the last two spectra of Figure \ref{spectroscopy}). 
We can use simple arguments to constraint the geometry of the system. Assuming a velocity of $\sim 10$ -{}- $100$ \kms~for the CS shell, and an initial radius of the massive shell of $R^{sh}_0=2.1\times 10^{14}$ cm, as suggested by the modelling, the inner shell must have been lost by the progenitor star $\sim 7$ -{}- $0.7$ years before the explosion. We can also compute when the outer CSM has been expelled by the progenitor through the formula:\\

$t_0 = [R^{sh}_0 + (v_{sh}-v_{CSM}) \times t_{int}]/v_{CSM}$,\\

\noindent
where $R^{sh}_0=2.1\times 10^{14}$ cm is the initial radius of the inner shell, $v_{sh} = 6300$ \kms~is the photospheric expansion velocity of the swept-up shell+ejecta, $v_{CSM} = 10$ -{}- $100$ \kms~is the assumed expansion for the CSM (similar to the expansion velocity of the massive shell before the collision with the ejecta), $t_{int} = 50$ days is the time after which the swept-up shell+ejecta starts to interact with the CSM (taking as reference the time of the first contact of the ejecta with the shell). The result is that the outer CSM has been lost by the progenitor star about $90$ -{}- $9$ years before the SN explosion.

At phases later than $+100$ days, the CSM surrounding the SN is almost swept-up and the radiation is coming from the inner part of the SN ejecta, however, as seen from the presence of the intermediate-width \Ha, some residual CSM interaction is still ongoing.\\
Because of the pronounced blueshift of the broad \Ha~emission seen at all phases, we suggest that the SN ejecta are asymmetric, with the asymmetric axes pointing in the observer's direction.

In summary, we propose that what we initially observe in \ass~is the result of the collision between the SN ejecta (not visible early on) with a pre-existing, massive shell. Since the late time spectra turn out to be very similar to the nebular spectra of SE-SN, we propose that the shell could be the result of mass loss due to a common envelope phase in a binary system. This is in fact the favoured channel proposed for stripping the external H-He mantle of massive stars before their explosion as SE SNe \citep{pod92,yoo10,eld13}. In binary systems, the SN progenitors may also be surrounded by dense CSM because of the  mass loss triggered by non-conservative mass transfer along the final phases of the progenitor evolution \citep{ouc17}.

Alternatively, a massive binary system composed by a H-poor LBV-like star that expelled a shell $\sim 7$ -{}- $0.7$ years before a Wolf-Rayet companion star exploded as \ass, could also explain the observations. A similar progenitor configuration has been proposed by \citet{pas07} to explain the output of the type Ibn SN~2006jc \citep[however, this scenario has been questioned by][]{mau16}.

\section{Conclusions} \label{conclusion}
Extensive photometric and spectroscopic observations of \ass~for slightly less than a year of its evolution are presented. The light curve show a slow evolution early on, which is followed by a faster evolution 100 days after maximum, with the $V$ light curve reaching an absolute  magnitude peak of $M_V \sim -19.6$.

The spectra up to $\sim 10$ days after maximum are featureless with BB temperature dropping from 13300 K at maximum to 8200 K 17 days later. At phase of $+11$ d, a He~I 5876 \AA~transition with a P-Cygni profile appears, with an expansion velocity of about 8000 \kms. At phases later than $+40$ days from the $V$ maximum, the blue region of the spectra are dominated by a blue pseudo-continuum frequently seen in interacting supernovae. Meanwhile, a broad \Ha~becomes more and more intense with time up to $\sim 150$ days after the $V$ band maximum and quickly declines later on. At the same phases an intermediate \Ha~component appears, which becomes narrower and fainter with time. The spectra at latest phases look very similar to the nebular spectra of stripped-envelope supernovae.

The early photometric evolution (the first 40 days after the V maximum) of the bolometric light curve is well reproduced by  a model in which the post-explosive evolution is dominated by the expanding ejecta and the energy budget is essentially coming from ejecta recombination and $^{56}$Ni decay. From this we infer, a mass of the ejecta, M$_{ej}=2.6$ \msun; an initial radius of the photosphere, R$_0 = 2.1\times 10^{14}$ cm; and an explosion energy, E$_{expl} = 0.8$ foe. 
The blue pseudo-continuum and the intermediate \Ha~component are, instead, clues that ejecta-circumstellar medium interaction starts from about 40 days after maximum. 

A possible scenario which accounts for all the observables involves a massive and extended H-poor shell lost by the progenitor star ($1-7$ years before explosion), probably during a common envelope episode with a companion. The shell is heated and accelerated by the collision with the ejecta and its emission models the early bolometric light curve. The accelerated shell+ejecta rapidly dilutes unveiling the unperturbed supernova spectrum below. Meanwhile, the shell+ejecta start to interacts with a more H-poor external circumstellar material lost by the progenitor system about $9-90$ years before the explosion. This interaction should occur in clumps or in highly asymmetric CSM, because only in this scenario, the inner ejecta are exposed.

This scenario involving the ejecta-massive shell collision could be a scaled down (both in mass and radius of the shell) version of those proposed for modelling some SLSNe.

\section*{acknowledgments}
We thank the ASAS-SN team for providing their photometry for ASASSN-15no. We also thank the anonymous referee for helpful comments.
SB, AP, NER, GT, MT and LT are partially supported by the PRIN-INAF 2014 (project `Transient Universe: unveiling new types of stellar explosions with PESSTO'). ST acknowledges support by TRR 33 "The Dark Universe" of the German Research Foundation (DFG). Based on observations made with: The Cima Ekar 1.82~m Telescopio Copernico of the Istituto Nazionale di Astrofisica of Padova, Italy. The Gran Telescopio Canarias (GTC) operated on the island of La Palma at the Spanish Observatorio del Roque de los Muchachos of the Instituto de Astrofisica de Canarias. The Nordic Optical Telescope (NOT), operated by the NOT Scientific Association at the Spanish Observatorio del Roque de los Muchachos of the Instituto de Astrofisica de Canarias. The Italian Telescopio Nazionale Galileo (TNG) operated on the island of La Palma by the Fundaci\'on Galileo Galilei of the INAF (Istituto Nazionale di Astrofisica) at the Spanish Observatorio del Roque de los Muchachos of the Instituto de Astrofisica de Canarias. The Telescopi Joan Or\`o of the Montsec Astronomical Observatory, which is owned by the Generalitat de Catalunya and operated by the Institute for Space Studies of Catalunya (IEEC). This work has made use of the NASA/IPAC Extragalactic Database (NED), which is operated by the Jet Propulsion Laboratory, California Institute of Technology, under contract with NASA. We also used NASA's Astrophysics Data System.

{}

\newpage
\begin{deluxetable*}{@{}crcccccl@{}}
\tablecaption{$ugriz$ photometry (AB system) of \ass.}
\tablehead{\colhead{MJD} & rest phase&\colhead{$u$(err)} & \colhead{$g$(err)} & \colhead{$r$(err)} & \colhead{$i$(err)} & \colhead{$z$(err)} & \colhead{Instrument} }
\startdata
57240.96 & +6& \nodata      & 16.36 (0.08) & 16.47 (0.06) & 16.59 (0.09) & \nodata      & AFOSC\\ 
57243.93 & +9& 16.65 (0.02) & 16.32 (0.05) & 16.44 (0.06) & 16.50 (0.08) & 16.52 (0.04) & AFOSC\\ 
57244.93 &+10& 16.62 (0.05) & 16.43 (0.04) & 16.48 (0.06) & 16.58 (0.06) & 16.62 (0.03) & AFOSC\\ 
57245.92 & +11& 16.69 (0.05) & 16.38 (0.05) & 16.48 (0.06) & 16.63 (0.10) & 16.63 (0.06) & AFOSC\\ 
57258.97 & +23& 17.28 (0.03) & 16.75 (0.05) & 16.63 (0.04) & 16.71 (0.04) & 16.70 (0.04) & ALFOSC-FASU\\ 
57263.91 & +28& \nodata      & \nodata      & 16.84 (0.19) & \nodata      & \nodata      & OSIRIS\\ 
57272.89 & +37& 17.50 (0.07) & 17.01 (0.07) & 16.93 (0.04) & 17.00 (0.08) & 16.78 (0.05) & AFOSC\\ 
57292.86 & +56& \nodata      & \nodata      & 17.18 (0.01) & \nodata      & \nodata      & OSIRIS\\ 
57309.87 & +72& 17.91 (0.06) & 17.54 (0.06) & 17.45 (0.07) & 17.45 (0.05) & \nodata      & LRS\\ 
57330.79 & +92& 18.12 (0.09) & 17.74 (0.07) & 17.54 (0.06) & 17.77 (0.10) & 17.27 (0.04) & AFOSC\\ 
57342.71 & +104& \nodata      & 17.89 (0.20) & 17.93 (0.08) & 18.07 (0.11) & 17.70 (0.06) & AFOSC\\ 
57359.22 & +120& 19.16 (0.17) & 18.34 (0.07) & 18.09 (0.06) & 18.26 (0.07) & 18.14 (0.05) & AFOSC\\ 
57386.22 & +146& 19.89 (0.14) & 18.88 (0.07) & 18.56 (0.06) & 18.74 (0.05) & 18.57 (0.07) & ALFOSC-FASU\\ 
57426.23 & +185& \nodata      & \nodata      & 19.84 (0.03) & \nodata      & \nodata      & OSIRIS\\ 
57429.19 & +187& $>$20.66 (0.36) & 20.09 (0.11) & 19.91 (0.11) & 20.13 (0.19) & 19.37 (0.09) & AFOSC\\ 
57464.19 & +221& $>$21.48 (0.51) & 21.08 (0.15) & 20.70 (0.13) & 20.95 (0.20) & 20.03 (0.20) & LRS\\ 
57479.18 & +236& \nodata         & \nodata      & 20.89 (0.50) & \nodata      & \nodata      & OSIRIS\\ 
57488.02 & +244& \nodata         & 21.48 (0.06) & 21.06 (0.10) & 21.23 (0.08) & \nodata      & LRS\\ 
57534.06 & +289& \nodata         & \nodata      & 21.59 (0.66)   & 21.53 (0.70) & 21.17 (0.70)& OSIRIS$^*$\\
\enddata
\tablecomments{$^*$ $i$ and $z$ measurements synthesised from the spectrum after normalising it to the $r$ photometric measurement. The rest phases are relative to the $V$ band maximum (2015 August 1).
}
\label{sdss_phot}
\end{deluxetable*}

\begin{deluxetable*}{@{}crcccccl@{}}
\tablecaption{$UBVRI$ photometry (Vega system) of \ass.}
\tablehead{\colhead{MJD} & rest phase& \colhead{$U$(err)} & \colhead{$B$(err)} & \colhead{$V$(err)} & \colhead{$R$(err)} & \colhead{$I$(err)} & \colhead{Instrument} }
\startdata
57233.31 & -2& \nodata      & \nodata      & $>$17.10     & \nodata      & \nodata      & ASASSN\\ 
57237.26 & +2& \nodata      & \nodata      & 16.46(0.09)        & \nodata      & \nodata      & ASASSN\\ 
57241.24 & +6& \nodata      & \nodata      & 16.50(0.13)        & \nodata      & \nodata      & ASASSN\\ 
57242.28 & +7& \nodata      & \nodata      & 16.62(0.09)        & \nodata      & \nodata      & ASASSN\\ 
57243.93 & +9& \nodata      & 16.55 (0.05) & 16.47 (0.05) & \nodata      & \nodata      & AFOSC\\ 
57244.28 & +9& \nodata      & \nodata      & 16.57(0.08)        & \nodata      & \nodata      & ASASSN\\ 
57244.92 & +10& \nodata      & 16.55 (0.06) & 16.54 (0.04) & \nodata      & \nodata      & AFOSC\\ 
57245.92 & +11& \nodata      & 16.70 (0.08) & 16.50 (0.05) & \nodata      & \nodata      & AFOSC\\
57246.26 & +11& \nodata      & \nodata      & 16.79(0.10)        & \nodata      & \nodata      & ASASSN\\ 
57250.26 & +15& \nodata      & \nodata      & 16.68(0.09)        & \nodata      & \nodata      & ASASSN\\ 
57255.25 & +20& \nodata      & \nodata      & 16.68(0.10)        & \nodata      & \nodata      & ASASSN\\ 
57258.96 & +23& \nodata      & 17.01 (0.03) & 16.69 (0.03) & \nodata      & \nodata      & ALFOSC-FASU\\ 
57316.80 & +79& 17.17 (0.11) & 17.65 (0.07) & 17.49 (0.06) & 17.22 (0.06) & 16.95 (0.05) & MEIA2\\ 
57317.80 & +80& 17.20 (0.12) & 17.71 (0.09) & 17.44 (0.05) & 17.36 (0.06) & 16.99 (0.07) & MEIA2\\ 
57330.79 & +92& \nodata      & 17.92 (0.06) & 17.63 (0.04) & \nodata      & \nodata      & AFOSC\\ 
57342.71 & +104& \nodata      & 18.19 (0.09) & 17.92 (0.09) & \nodata      & \nodata      & AFOSC\\ 
57359.23 & +120& \nodata      & 18.57 (0.10) & 18.20 (0.05) & \nodata      & \nodata      & AFOSC\\ 
57386.21 & +146& \nodata      & 19.05 (0.07) & 18.80 (0.07) & \nodata      & \nodata      & ALFOSC-FASU\\ 
57429.20 & +187& \nodata      & 20.10 (0.15) & 19.88 (0.19) & \nodata      & \nodata      & AFOSC\\ 
57488.04 & +244& \nodata      & 21.76 (0.18) & 21.38 (0.19) & \nodata      & \nodata      & LRS\\ 
\enddata
\tablecomments{The rest phases are relative to the $V$ band maximum (2015 August 1).
}
\label{vega_phot}
\end{deluxetable*}

\end{document}